\documentstyle[11pt,psfig,epsf]{article}
\def\reference{\parskip 0pt\par\noindent\hangindent 0.5 truecm}

\def\simlt{\lower.5ex\hbox{$\, \buildrel < \over \sim \,$}}
\def\simgt{\lower.5ex\hbox{$\, \buildrel > \over \sim \,$}}

\begin{document}
\title{\bf TTF: A Flexible Approach to Narrowband Imaging}

\author{Joss Bland-Hawthorn$^1$ \& Heath Jones$^2$}
\date{}
\maketitle

{\center
$^1$ Anglo-Australian Observatory, P.O. Box 296, Epping, NSW 2121
\\jbh@aaossz.aao.gov.au\\[3mm]
$^2$ Mount Stromlo \& Siding Spring Observatories,\\ Private Bag,
Weston Creek P.O.,\ ACT 2611\\dhj@mso.anu.edu.au\\[3mm]
}

\begin{abstract}
The Taurus Tunable Filter ({\sc TTF}) is a tunable narrowband interference
filter covering wavelengths from 6300\AA\ to the sensitivity
drop-off of conventional CCDs ($\sim9600$ \AA), although a blue `arm' 
(3700$-$6500\AA) is to be added by the end of 1997. 
The TTF offers monochromatic imaging
at the cassegrain foci of both the Anglo-Australian and William Herschel 
Telescopes, with an adjustable passband of between 6 and 60 \AA. In addition,
frequency switching with the TTF can be synchronized to movement of charge
(charge shuffling) on the CCD which has important applications to many 
astrophysical problems. Here we review the different modes of TTF and suggest
their use for follow-up narrowband imaging to the AAO/UKST Galactic Plane 
H$\alpha$ Survey. 
\end{abstract}

{\bf Keywords: instrumentation: detectors --- methods: observational\\
~~~--- techniques: interferometric}

\bigskip

\section{INTRODUCTION}

The Taurus Tunable Filter (TTF), manufactured by Queensgate Instruments
Pty.\ Ltd., has the appearance of a conventional Fabry-Perot etalon
in that it comprises two highly polished glass plates.
Unlike conventional Queensgate etalons, however, the TTF incorporates very 
large piezo-electric stacks (which determine the plate separation) and
high performance coatings over half the optical wavelength range. Moreover,
the plate separation can be tuned to spacings as small as 2$\mu$m 
(measured at the coating surfaces).
The TTF is used in the collimated beam of the TAURUS-2 focal reducer 
available at both the 3.9 m Anglo-Australian (AAT) and 4.2 m William Herschel 
(WHT) telescopes. Field coverage is 10 arcmin at f/8 or 5 arcmin at f/15.

For the first time, TTF provides the capability to synchronize
frequency switching with the movement of charge on a CCD, or
charge shuffling. This has important benefits for many astrophysical
experiments, not least for averaging out temporal variations due to the
atmosphere or measurement apparatus. This instrument is an important step
in changing the way that intermediate to narrowband imaging is
performed at observatories.

The TTF has largely removed the need for buying arbitrary narrow and
intermediate interference filters, as one can tune the bandpass and
the centroid of the bandpass by selecting the plate spacing. 
The spacing of the plates is controlled to extremely high
accuracy with a capacitance bridge (Jones \& Richards 1968). 
This approach to tunable imaging has existed since the instrument of 
Atherton \& Reay (1981), although TTF is the first of its
kind in terms of both wavelength and bandpass accessibility.
Since tunable filters have a periodic transmission profile, the instrument
requires a limited number of blocking filters. At low resolution ($R = 300$),
conventional broadband $R$ and $I$ filters suffice. At high resolution
($R = 1000$), eight intermediate band filters are used to
sub-divide $R$ and $I$.

The highly polished plates are coated for optimal performance over 
6300-9600 \AA. The coating reflectivity (96 \%) determines the
shape and degree of order separation of the instrumental profile. This
is fully specified by the coating finesse, $N$, which has a
quadratic dependence on the coating reflectivity. The TTF was coated
to a finesse specification of $N = 40$ which means that the separation
between periodic profiles is forty times the width of the
instrumental profile. At such high values, the profile is Lorentzian
to a good approximation. For a given wavelength, changes in plate
spacing, $L$, correspond to different orders of interference, $m$. This in
turn, dictates the resolving power $R = Nm$ according to the finesse.

In the following sections we summarise the different observing modes
of TTF. Discussion is also made of phase effects
in the field and their influence. The flexibility of
TTF will see it well-suited to narrowband follow-up from the AAO/UKST
Galactic Plane H$\alpha$ Survey, in lines such as H$\alpha$, 
[NII]$\lambda$6583, [SII]$\lambda$6717 and [SII]$\lambda$6731. 

We maintain a WWW site ({\tt http://msowww.anu.edu.au/$\sim$dhj/ttf.html})
describing all aspects of TTF and its operation.
In addition to general use, the instrument is available for AAT
service time ({\tt http://www.aao.gov.au/local/www/jmc/service/service.html})
if observations are shorter than 3 hours.

\section{CCD CHARGE SHUFFLING}

\begin{figure}[h]
\centerline{\psfig{file=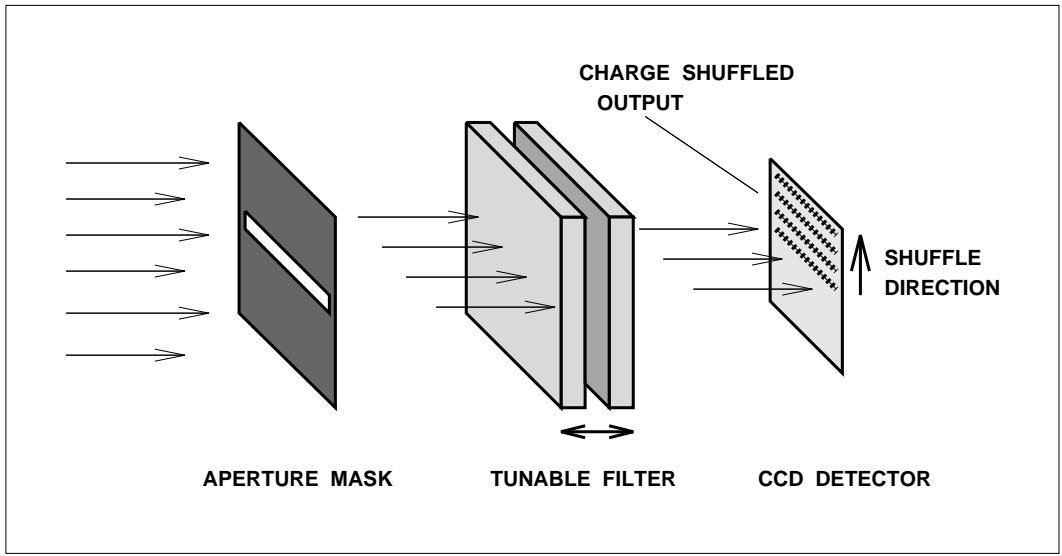,width=140mm}}
\caption{Simplified view of how the CCD chip is
illuminated and shuffled. The mask aperture size controls the amount
of the CCD exposed and therefore the number of different images fit on
to an individual frame. Charge can be shuffled in both up and and down
the chip.}
\label{longslit}
\end{figure}

Central to almost all modes of TTF use is charge shuffling.
Charge shuffling is movement of charge along the CCD
between multiple exposures of the same frame, before the image
is read out (Clemens \& Leach 1987, Cuillandre et al.\ 1994).
An aperture mask ensures only a section of the CCD frame is
exposed at a time. For each exposure, the tunable filter is systematically 
moved to different gap spacings in a process called frequency-switching.
This way, a region of sky can be captured at several different
wavelengths on the one image (Fig.\ \ref{longslit}).
Alternatively, the TTF can be kept at fixed frequency and
charge shuffling performed to produce time-series exposures.
Each of these modes is described in the following section.

The TTF plates can be switched anywhere over the physical range
2 to 12 $\mu$m at rates in excess of 100 Hz, although in most
applications, these rates rarely exceed 0.1 Hz. If a shutter is used,
this limits the switching rate to about 1 Hz. Charge on a large format
CCD can be moved over the full area at rates approaching 10 Hz: it is
only when the charge is read out through the amplifiers that this rate
is greatly slowed down. The TTF exploits the ability of certain large
format CCDs to move charge up and down many times before significant
signal degradation occurs (Yang et al.\ 1989). 
In this way, it is possible to form
discrete images taken at different frequencies where each area of the
detector may have been shuffled into view many times to average out
temporal effects.

The field of view available in shuffle mode depends on the number of
frequencies being observed. When we move the charge upwards, say,
information in pixels at the top of the field is rolled off the top
and lost.  For example, two frequencies requires that we divide the
CCD into three vertical partitions where information in one of the
outer partitions is lost in the shuffle process. In the limit of $n$
frequencies, where $n$ is large, only half the available detector area
is used to store information. The new MIT-LL 4096 x 2048
(rows $\times$ columns) format CCDs will increase the detector area
available for shuffling by a factor of 4 compared to the present Tek
1024$\times$1024 CCD. This is because the instrument field of view
projects to an aperture 1024 pixels in diameter and shuffling is only
possible in the vertical direction. 

One application of charge shuffling
that does not sacrifice detector area is time-series imaging (Section 3.3,
below). This is because the imaging region is relatively small and able to
be read out as the time-series progresses.

\section{OBSERVING MODES OF TTF}

There are several technical problems driving the development
of tunable filters for narrowband imaging over standard fixed
interference filters. First, it is difficult and very expensive for
manufacturers to produce high performance narrow passbands, 
particularly at resolving powers approaching 1000. This problem is
largely circumvented by use of TTF in conjunction with a 5-cavity,
intermediate bandpass blocking filter. Secondly, the TTF instrumental
function has identical form at all wavelengths and all bandpasses. A
comparison of two discrete bands at different wavelengths is moderated
only by the blocking filter transmission at those wavelengths (which
is normally flat in any case), and the ratio of the gap spacings.
Thirdly, the same optical path is used at all frequencies.
Furthermore, the ability to shuffle charge allows one to average out
all temporal variations: atmospheric transparency, the contribution
from atmospheric lines, seeing, detector and electronic instabilities.

We now describe some of the advantages to be had from the TTF
shuffle capability in imaging emission-line sources.

\subsection{Tuning to a Specific Wavelength and Bandpass}
\begin{figure}[h]
{\LARGE (a)}
\psfig{file=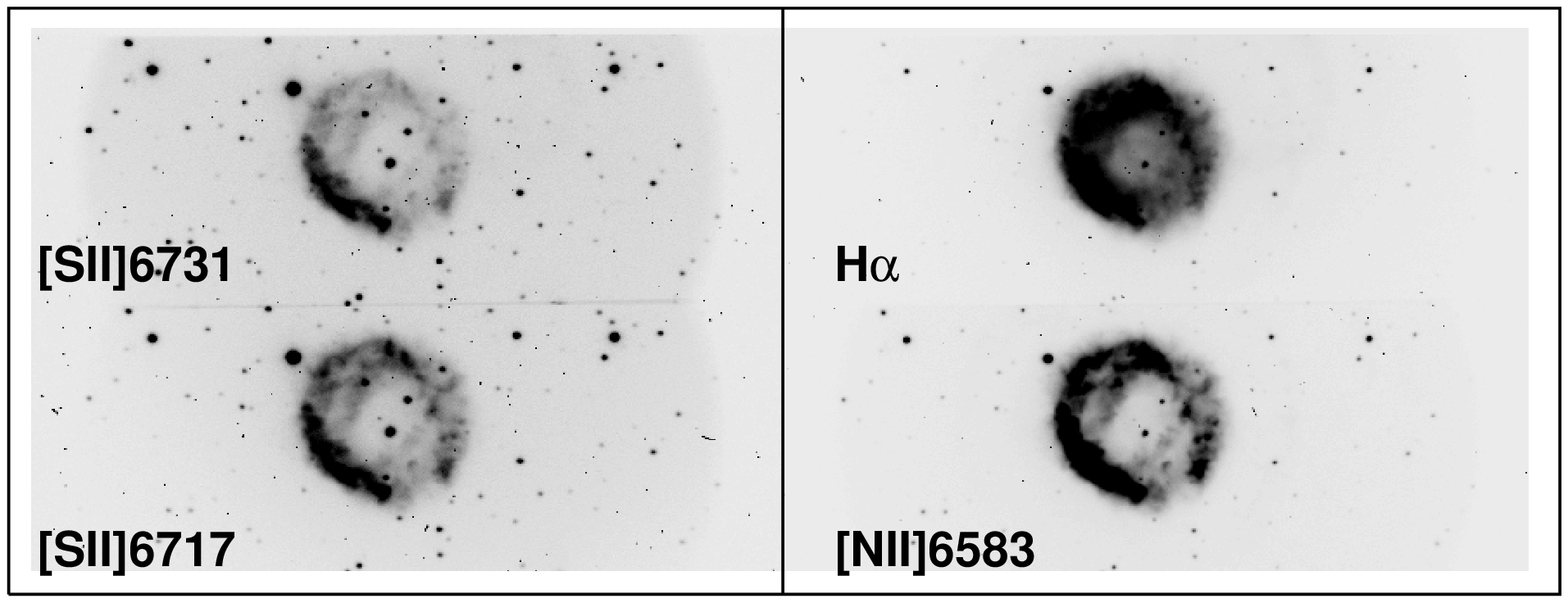,height=70mm,bbllx=18pt,bblly=18pt,bburx=274pt,bbury=208pt,clip=}
{\LARGE (b)}
\psfig{file=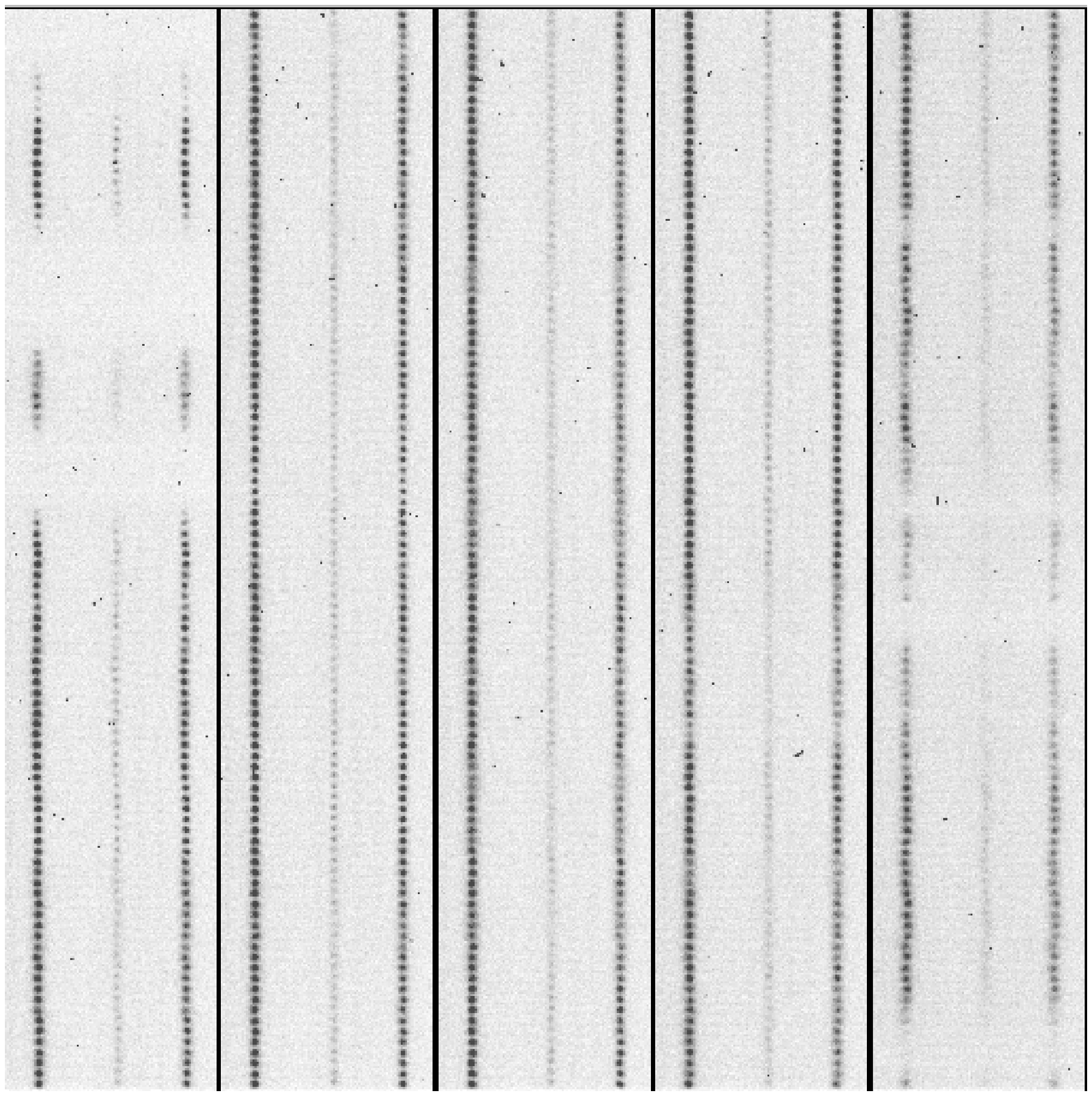,height=70mm,bbllx=148pt,bblly=140pt,bburx=255pt,bbury=355pt,clip=}\\
{\LARGE(c)}
\psfig{file=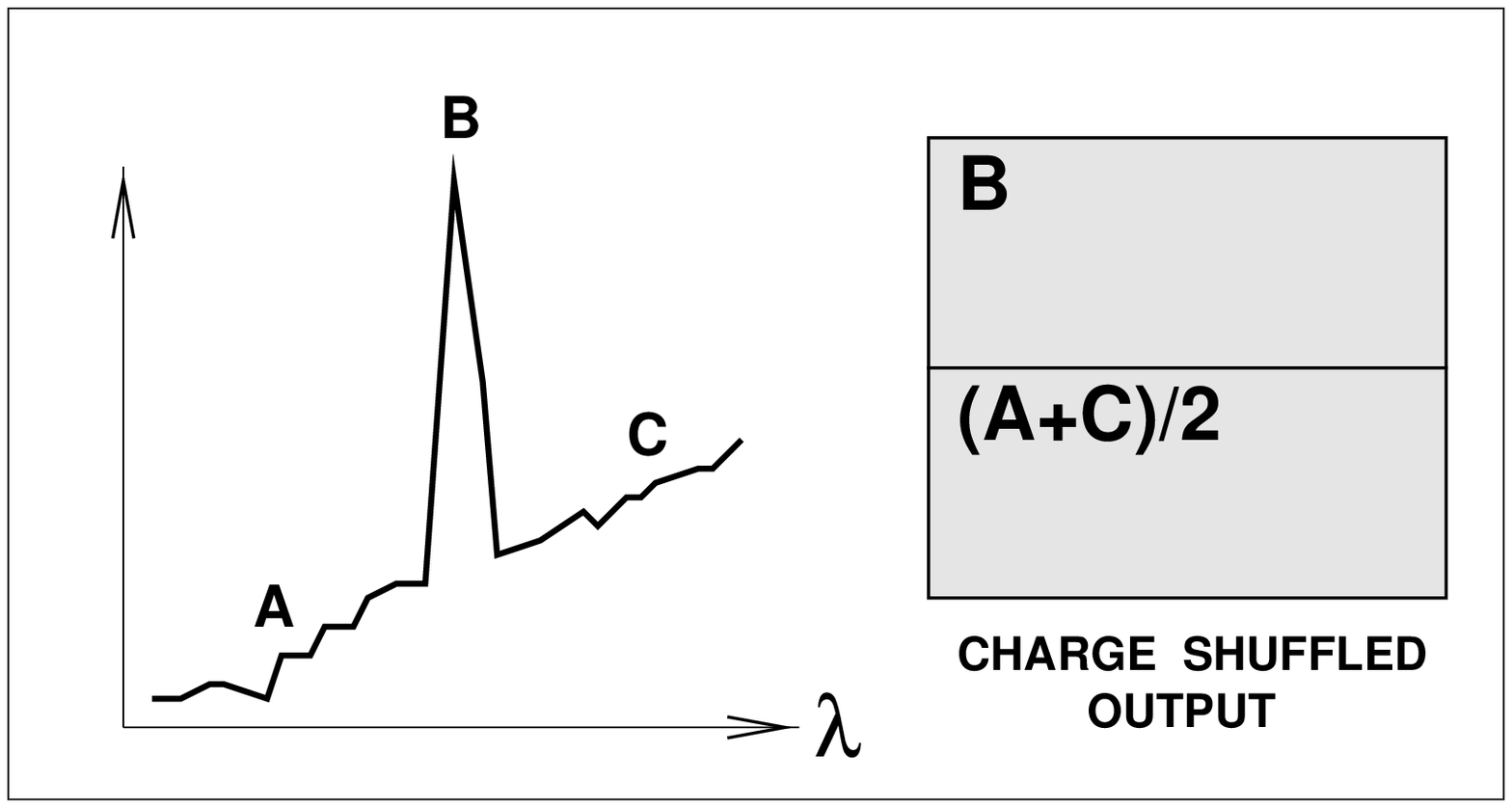,width=150mm,bbllx=67pt,bblly=269pt,bburx=542pt,bbury=540pt,clip=}

\caption{($a$) Charge shuffled image of the planetary nebula NGC 2438
in [SII]$\lambda$6371 ({\em top}) and [SII]$\lambda$6717 ({\em bottom}).
The TTF was frequency-switched 18 times during the 12 min exposure and the
charge shuffled back and forth in concert. A 10 \AA\ bandpass was used.
($b$) Charge shuffle time-series imaging of V2116 Oph, an x-ray pulsar
in orbit around a red giant. V2116 Oph is to the right, reference stars 
are at the centre and left. TTF was tuned to a 7 \AA\ bandpass
centred on the [OI]$\lambda$8446 line, which is suspected to vary 
with the pulsar period of 120 s. Each exposure was 12 s.
($c$) Charge shuffling either side of an emission-line to average out
the effect of a steeply sloping continuum. Equal time is spent
collecting continuum from both sides of the emission line. }
\label{shuffExamples}
\end{figure}

This allows us to obtain images of obscure spectral lines at arbitrary
redshifts (see e.g.\ Jones \& Bland-Hawthorn 1997a). 
We can also optimize the bandpass to accomodate the line
dispersion and to suppress the sky background. The off-band frequency
can be chosen to avoid night-sky lines and can be much wider so that
only a fraction of the time is spent on the off-band image.
Fig.\ \ref{shuffExamples}($a$) shows a charge-shuffled image of 
the planetary nebula NGC 2438 in [SII] lines at 6731 and 6717 \AA. 
During this 12 min exposure, TTF was switched between the two frequencies
18 times while the charge was shuffled back and forth accordingly.
In this way, temporal variations in atmospheric transmission are
equally shared between each passband over the entire exposure time.

\subsection{Shuffling Between On and Off-band Frequencies}
With the new MIT-LL chips, we are able to image the full 10 arcmin 
field for two discrete frequencies. We can also choose a narrow
bandpass for the on-band line and a much broader bandpass (factor of 4-5)
for the off-band image so that we incur only a 20 \% overhead for the
off-band image. As with specific tuning (Section 3.1, above),
multiple frequencies can be imaged in a single frame, the number of
which is entirely arbitrary.

\subsection{Time-Series Imaging} 
For time-varying sources, we can step the charge in one direction while
only switching between a line and a reference frequency. For
example, a compact variable source imaged through a narrow aperture in
the focal plane forms a narrow image at the detector. We can switch
between the line and a reference frequency many times forming a set of
narrow interleaved images at the detector. The reference frequency is
a measure of the atmospheric stability during the time series.
Alternatively, if nearby
reference stars are available for photometric calibration, then charge
shuffling can be done at a single fixed frequency as demonstrated in 
Fig \ref{shuffExamples}($b$). For example, some x-ray
binaries produce strong emission lines that vary on 0.1 Hz
timescales.

An important difference between time-series and multi-frequency shuffles
is that the former does {\em not} sacrifice any detector area. This 
is possible only when charge shuffling occurs in the same direction as
normal CCD read out. For example, with a slit only 4 pixels wide, we
can obtain 500 images in the emission line, interleaved with a further
500 images at a reference frequency.

\subsection{Adaptive Frequency Switching with Charge Shuffling}
When imaging spectral lines that fall between OH bandheads or on steeply
rising continua, a powerful feature is the ability to shuffle between on 
and off bands but where the off band is alternated between two or more
frequencies either side of the on-band frequency. As illustrated in
Fig.\ \ref{shuffExamples}, this can be used to average out either
rapid variations in blocking filters or the underlying spectral continuum.

\subsection{Shuffling Charge in One Direction with a Narrow Focal Plane Slit}
This allows us to produce a long-slit spectrum of an extended
source. While this is vastly less efficient than using a long-slit
spectrograph, this is a fundamental requirement for establishing the
parallelism of reflecting mirrors at few micron spacings (Jones \& 
Bland-Hawthorn 1997b). Conventional methods are an order of magnitude slower.

\section{PHASE EFFECTS}

\begin{figure}[h]
 \centerline{\epsfxsize=100mm\epsfbox{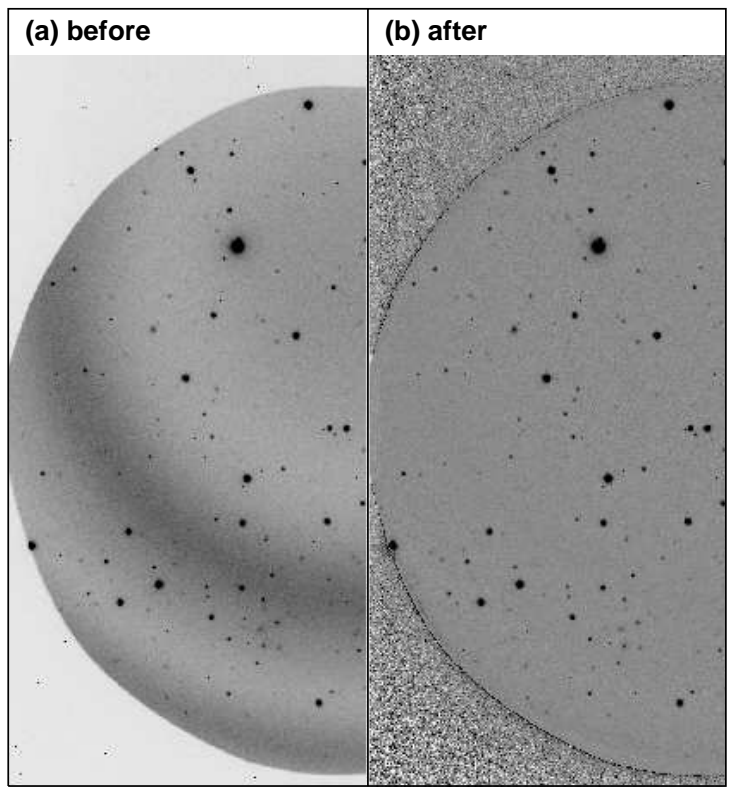}}
\caption{Images showing the removal of the atmospheric OH emission lines
from a raw TTF image before ($a$) and after ($b$) cleaning. Only half the
full TTF field is shown in each case. }
\label{wavgrad}
\end{figure}

The primary goal of a tunable filter is to provide a monochromatic
field over as large a detector area as possible. With the present TTF,
however, the field of view is not strictly monochromatic. 
The effect is most acute at high orders of interference. 
Fig.\ \ref{wavgrad}($a$) shows how wavelength 
gradients (or phase effects) are evident from a ring pattern of 
atmospheric OH emission lines across the TTF field.

Wavelengths are longest at the centre and get bluer the further
one moves off-axis. For instruments such as TTF, the wavelength as a
function of off-axis angle $\theta$ is
\begin{equation}
\lambda (\theta) = \lambda_{\rm centre} \cdot \Bigl ( 1 - \frac{\theta^2}{2}
\Bigr ).
\label{eqn6}
\end{equation}
Here $\lambda_{\rm centre}$ is the on-axis wavelength, equal to $2 L/m$
(for an air gap)\footnote{
Eqn.\ (\ref{eqn6}) makes use of the small angle
approximation for $\cos \theta$.},
where $L$ is the physical plate spacing and $m$ is the order of interference.

It follows from Eqn.\ (\ref{eqn6}) that the change in wavelength
$\lambda_\phi$ across an angle $\theta$ from the centre is given by
\begin{equation}
\lambda_\phi (\theta) =  \lambda_{\rm centre} \cdot \frac{\theta^2}{2}.
\label{eqn8}
\end{equation}
Since $\theta \leq 0.14$ for the current Tektronix CCD, then
$\lambda_\phi / \lambda_{\rm centre}$ is {\em always}
$\simlt 10^{-2}$. It also follows that $\lambda_\phi / \lambda_{\rm centre}$
remains fixed over a given radius, irrespective of order $m$.
Bland \& Tully (1989) derive similar equations in the context
of a higher order conventional etalon.

We define our monochromatic field by the size of the Jacquinot spot,
the central region of the ring pattern. By definition, the Jacquinot spot is
the region over which the wavelength changes by no more than $\sqrt{2}$ of
the etalon bandpass, $\delta\lambda$. For wavelength, $\lambda$, the 
bandpass relates directly to the order $m$ such that 
\begin{equation}
\delta\lambda = \lambda / Nm = \lambda / (40 \times m).
\label{eqn8.5}
\end{equation}
Here, $N$ is the effective finesse of the etalon, which in the case
of TTF is approximately 40. Combining Eqns.\ (\ref{eqn6}) and (\ref{eqn8.5})
we find that the angle subtended by the Jacquinot spot is
\begin{equation}
{\theta_{\rm Jac}}^2 = 2 \sqrt{2}/ N m  {\rm \, \,} = 1/(14.1 \times m) .
\label{eqn9}
\end{equation}
For a particular etalon, the size of the Jacquinot spot depends on
order $m$ alone. Eqn.\ (\ref{eqn9})
shows how the spot covers increasingly larger areas on the detector
as the filter is used at lower orders order of interference. The absolute
wavelength change across the detector remains the same, independently of
order. However, its effect relative to the bandpass diminishes as $m$
decreases. Fig.\ \ref{wavgrad}($b$) demonstrates how the effects of
atmospheric emission can be removed during reduction. 

The TTF is the most straightforward application of tunable filter
technology. Other, more sophisticated techniques such as
acousto-optic filters exist, (see Bland-Hawthorn \& Cecil 1996 for a review),
although all are currently considerably more expensive.
In future TTF-type instruments, phase effects will be 
eliminated from the outset by bowed plates. One advantage of such a
design is that the TTF will no longer have to be tilted to deflect ghost
reflections. Furthermore, interference coatings are notorious for bowing
plates and this can be factored into the plate curvature
specification. Other possible improvements are additional cavities
to square up the instrument profile. All of these modifications are
currently being explored.

\section{SUMMARY}

We have discussed details of the Taurus Tunable Filter (TTF) instrument and
its use. When used in conjunction with a CCD charge shuffling technique, 
TTF is well-suited for multi-narrowband imaging of emission-line sources.  
As such, it is ideal for follow-up imaging to the AAO/UKST
Galactic Plane H$\alpha$ Survey, for sources that are $\sim 10$ arcmin or
smaller. The analysis of tunable filter data is more straightforward than 
traditional Fabry-Perot imaging as all regions of the field are close to a
common wavelength.

The present TTF instrument exhibits some phase effects at high resolving
powers. We have demonstrated that these effects are tolerable for most
applications ($\lambda_\phi / \lambda_{\rm centre} \simlt 10^{-2}$) 
and the side-effects correctable through software. The commissioning of
TTF has marked the start of an exciting period of new imaging instruments 
for optical astronomy.

\section*{Acknowledgements}

Thanks to J.R. Barton, L.G. Waller, T.J. Farrell, E.J. Penny and C. McCowage
for technical input during TTF and charge-shuffle implementation. 
DHJ acknowledges the assistance of a Commonwealth Australian Postgraduate
Research Award.

\section*{References}

\reference Atherton, P.~D.~and Reay, N.~K.~1981 MNRAS 197, 507

\reference Bland, J.~and Tully, R.~B.~1989 AJ 98(2), 723

\reference Bland-Hawthorn, J.~and Cecil, G.~N.~1996 in Atomic, Molecular
and Optical Physics: Atoms and Molecules, vol.~29B, ch.~18, 
eds.\ Dunning, F.~B.~and Hulet, R.~G., Academic Press, ({\tt astro-ph/9704142})

\reference Clemens, D.~P.~and Leach, R.~W.~1987 Opt.~Eng.~26, 9

\reference Cuillandre, J.~C., Fort, B., Picat, J.~P., Soucail, G., Altieri, B., Beigbeder, F., Dupin, J.~P., Pourthie, T.~and Ratier, G.~1994, Astron.~Astrophys.~281, 603

\reference Jones, H.~and Bland-Hawthorn, J.~1997a PASA, 14, 8

\reference Jones, D.~H.~and Bland-Hawthorn, J.~1997b Appl.~Opt., in prep.

\reference Jones, R.~V.~and Richards, J.~C.~S.~1973 J.~Phys.~E.~6, 589

\reference Yang, F.~H., Blouke, M.~M., Heidtmann, D.~L., Corrie, B.,
Riley, L.~D.~ and Marsh, H.~H.~1989 in Optical Sensors and
Electronic Photography (Proc.~SPIE), eds.\ Blouke, M.~M.~and Pophal, D.,
SPIE -- International Society for Optical Engineering, 213

\newpage

\newpage

\newpage

\end{document}